\documentclass{webofc}
\usepackage[varg]{txfonts}   
\usepackage{bm}
\usepackage{amssymb}

\newcommand{\be}{\begin{equation}}
\newcommand{\ee}{\end{equation}}
\newcommand{\bea}{\begin{eqnarray}}
\newcommand{\eea}{\end{eqnarray}}
\newcommand{\lan}{\langle}
\newcommand{\ran}{\rangle}

\begin{document}
\title{Formal Theory of Heavy Ion Double Charge Exchange Reactions}

\author{\firstname{Horst} \lastname{Lenske}\inst{1,5}\fnsep\thanks{\email{horst.lenske@physik.uni-giessen.de}}
\and
\firstname{Jessica} \lastname{Bellone}\inst{2,5} 
\and
\firstname{Maria} \lastname{Colonna}\inst{2,3,5} 
\and
\firstname{Danilo} \lastname{Gambacurta}\inst{2,3,5} 
\and
\firstname{Jose-Antonio} \lastname{Lay}$^{4,5}$  
}

\institute{Institut f\"ur Theoretische Physik, JLU Giessen, D-35392 Giessen, Germany
\and
Istituto Nazionale di Fisica Nucleare, Laboratori Nazionali del Sud, Catania, Italy
\and
Dipartimento di Fisica e Astronomia "E. Majorana", Universit\`{a} di Catania, Catania, Italy
\and
University of Sevilla, Sevilla, Spain
\and
The NUMEN Collaboration, LNS Catania, I-95123 Catania, Italy
}
\abstract{
The theory of heavy ion double charge exchange (DCE) reactions $A(Z,N)\to A(Z\pm 2,N\mp 2)$ is recapitulated emphasizing the role of Double Single
Charge Exchange (DSCE) and pion-nucleon Majorana DCE (MDCE) reactions. DSCE reactions are of second--order distorted wave character, mediated by isovector nucleon-nucleon (NN) interactions. The DSCE response functions resemble the nuclear matrix elements (NME) of $2\nu 2\beta$  decay.
The MDCE process proceeds by a dynamically
generated effective rank-2 isotensor interaction, defined by off--shell pion--nucleon DCE scattering. In closure approximation pion potentials and two--nucleon correlations are obtained, similar to the neutrino potentials and the intranuclear exchange of Majorana neutrinos in $0\nu 2 \beta$ Majorana double beta decay (MDBD).
}
\maketitle
\section{Introduction}\label{intro}
For a long time, heavy ion charge exchange reactions were considered as determined by transfer processes in which protons and neutrons were reshuffled in sequences of nucleon or nucleon cluster exchange processes. In the 1980ies, however, the increasing amount of data on heavy ion single charge exchange (SCE) reactions led to growing evidence that the mean-field driven transfer processes were not describing adequately the measured cross sections. The search for alternative reaction scenarios led to the conclusion that collisional nucleon--nucleon (NN) processes need to be considered. Collisional SCE is mediated by the isovector NN interactions from the exchange of (virtual) $\pi$--, $\rho$--, and $\delta/a_0(980)$--mesons, respectively, including rank--0 central and rank--2 tensor spin--spin interaction, \cite{Lenske:2018jav,Lenske:2019cex}.

As discussed in depth in Ref. \cite{Cappuzzello:2022ton} the understanding of DCE reactions followed a similar scheme. The early hypothesis that sequential proton and neutron pair transfers dominate were questioned quite recently by explicit investigations showing that pair transfer is suppressed especially in mass-asymmetric systems \cite{Cappuzzello:2022ton} and strong contributions of second--order  mesonic DCE reaction mechanisms prevail. In this contribution, two competing and interfering reaction mechanisms relying on  collisional interactions are considered: The nucleon-nucleon Double Single Charge Exchange (DSCE) and meson-nucleon Majorana DCE (MDCE) are reviewed in brief in the following sections.

\section{Double Single Charge Exchange Reactions}\label{sec:DSCE}

DSCE reactions are a sequence of two single charge exchange events, each of them mediated by the two--body NN--isovector interaction $\mathcal{T}_{NN}$, the latter acting by one--body operators on the projectile and the target nucleus, respectively. For a reaction $\alpha= a(Z_a,N_a)+A(Z_A,N_A) \to \beta=b(Z_a\pm 2,N_a\mp 2)+B(Z_A\mp 2,N_A\pm 2)$ the reaction amplitude is written down readily as a quantum mechanical second order distorted wave reaction matrix element \cite{Bellone:2020lal}:
\be\label{eq:MDSCE}
\mathcal{M}^{(2)}_{\alpha\beta}(\mathbf{k}_\alpha,\mathbf{k}_{\beta})=\lan \chi^{(-)}_\beta, bB|\mathcal{T}_{NN}\mathcal{G}^{(+)}_{aA}(\omega_\alpha)\mathcal{T}_{NN}|aA,\chi^{(+)}_{\alpha} \ran  .
\ee
Initial (ISI) and final state (FSI) interactions are taken into account by the distorted waves $\chi^{(\pm)}_{\alpha,\beta}$, depending on the center--of--mass (c.m.) momenta $\mathbf{k}_{\alpha,\beta}$ and obeying outgoing and incoming spherical wave boundary conditions, respectively. The available c.m. energy is $\omega_\alpha=\sqrt{s_{aA}}$ where $s_{aA}=(T_{lab}+M_a+M_A)^2-T_{lab}(T_{lab}+2M_a)$ if $a$ is the incoming nucleus.

As discussed in \cite{Lenske:2018jav}, we use an (anti--symmetrized and complex--valued) isovector NN T-matrix. For DCE reactions the isospin--changing parts of the NN T-matrix are of relevance, $\mathcal{T}_{NN}(i,j)=T_{NN}(i,j)\tau^{\pm}_i\tau^{\mp}_j$, $i \in \{a\}$ and $j \in \{A\}$. The reduced T--matrix is a superposition of rank--0 spin--scalar and spin--vector central interactions,
$T_{ST}=V_{ST}\left[\bm{\sigma}_a\cdot\bm{\sigma}_A\right]^S$, and rank--2 spin--tensor interactions,
$T_{Tn}=V_{Tn}Y_2\cdot \left[\bm{\sigma}_a\otimes\bm{\sigma}_A\right]_2$, respectively, where $S=0,1$ and $T=1$. All operators are contracted to total scalars.

\begin{figure}
\begin{center}
\sidecaption
\includegraphics[width = 6.5cm]{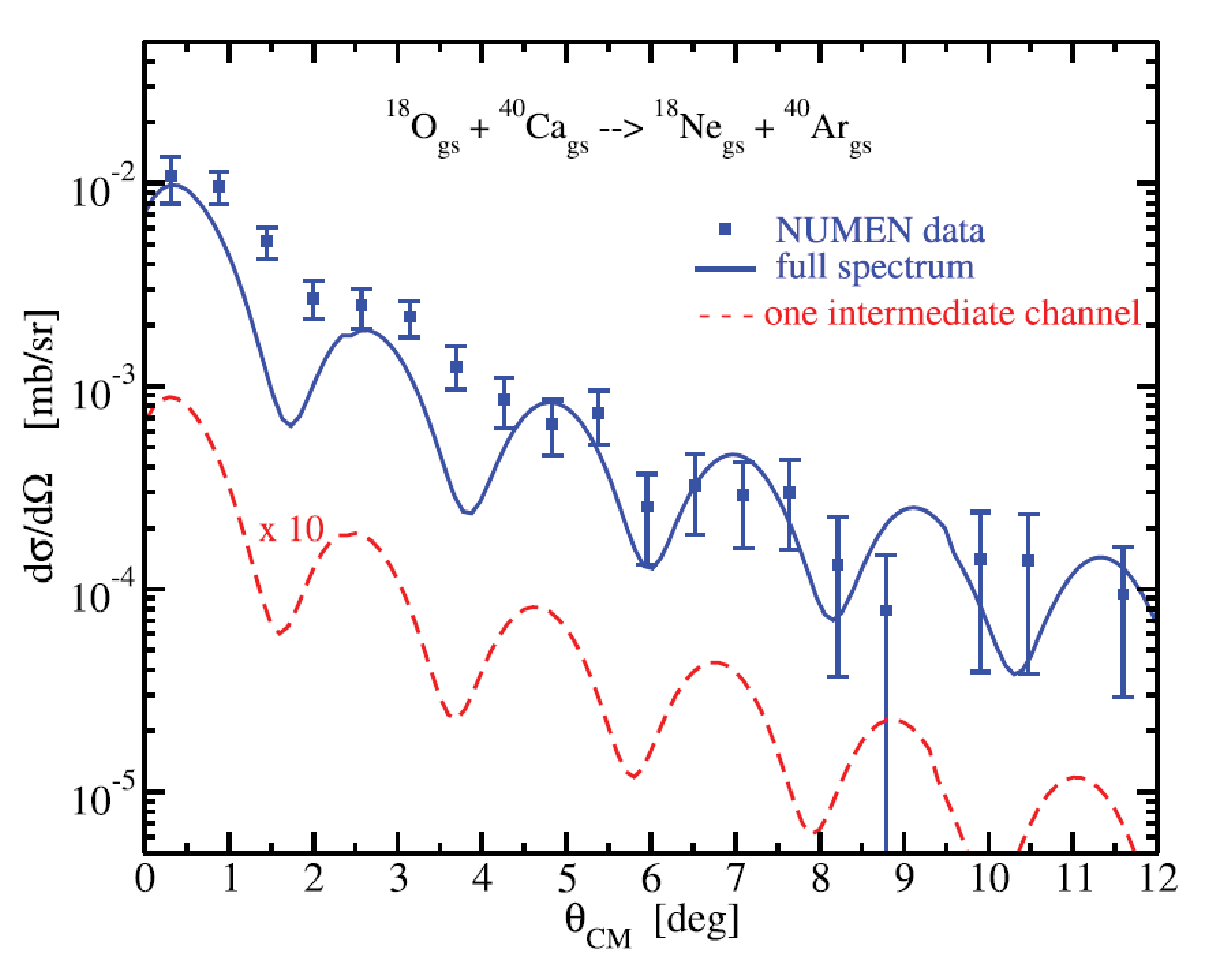}
\caption{DCE angular distribution for the reaction $^{40}$Ca $(^{18}$O, $^{18}$Ne$_{g.s.})^{40}$Ar$_{g.s.}$ at $T_{lab}=275$~MeV. Theoretical DSCE  results are compared to the NUMEN data \cite{Cappuzzello:2015ixp}. Intermediate states with angular momenta $J^\pi=5^\pm$ are included. The theoretical angular distribution is folded with the experimental angular resolution of $\pm 0.6^\circ$ (from \cite{Bellone:2020lal}). For comparison, the DSCE cross section obtained by going though only the lowest $J^\pi=1^+$ state in $^{40}$K is shown (dashed line).}
\label{fig:dSCE_xsec_PLB}
\end{center}
\end{figure}

With the set of intermediate SCE--type states $\{|c\ran\}$ and $\{|C\ran\}$ in projectile and target, respectively, the reaction amplitude attains the standard form of second order perturbation theory \cite{Lenske:2018jav,Bellone:2020lal}
\be\label{eq:MDSCE1}
\mathcal{M}^{(2)}_{\alpha\beta}(\mathbf{k}_\alpha,\mathbf{k}_\beta)=\sum_{\gamma=\{c,C\}}
\int \frac{d^3k_\gamma}{(2\pi)^3}M^{(1)}_{\gamma\beta}(\mathbf{k}_\gamma,\mathbf{k}_\beta)
\frac{\tilde{S}^\dag_{\gamma}}{\omega_\alpha-M^*_c-M^*_C-T_\gamma(k_\gamma)+i\eta}
M^{(1)}_{\alpha\gamma}(\mathbf{k}_\alpha,\mathbf{k}_\gamma).
\ee
$M^*_{c,C}=M_{c,C}+\varepsilon_{c,C}$ are the total rest masses of the intermediate nuclei including excitation energies $\varepsilon_{c,C}$, respectively. $T_\gamma$ denotes the kinetic energy related to the (off--shell) momentum $k_\gamma$. $M^{(1)}_{\alpha\gamma}(\mathbf{k}_\gamma,\mathbf{k}_\alpha)$ are the -- half off--shell -- SCE--amplitudes. $\tilde{S}^\dag_\gamma\sim \lan \tilde{\chi}^{(+)}_\gamma|\tilde{\chi}^{(-)}_\gamma\ran$ is the dual S--matrix element related to the non--hermitian Hamiltonian of relative motion \cite{Lenske:2021bpk,Lenske:2019cex,Bellone:2020lal}.

In Fig. \ref{fig:dSCE_xsec_PLB} the theoretical DSCE \cite{Bellone:2020lal} cross section is compared to the measured DCE angular distributions for the reaction $^{40}$Ca $(^{18}$O, $^{18}$Ne$)^{40}$Ar at $T_{lab}=275$~MeV \cite{Cappuzzello:2015ixp}. The theoretical results were
obtained by evaluating Eq.\eqref{eq:MDSCE1} numerically in pole approximation. The transitions in the nuclei involved in the reaction of
Fig.\ref{fig:dSCE_xsec_PLB} were describe in Quasiparticle Random Phase Approximation (QRPA), using the spectral distributions discussed in  \cite{Lenske:2018jav}.
A large spectrum of intermediate states up to $J^\pi=5^\pm$ was included. The magnitude of the measured cross section is rather well reproduced without the need of scaling factors. The calculations show, however, a more pronounced diffraction pattern than observed, leaving room for further contributions.

As seen above, DSCE reaction theory is based dynamically on a ladder-type t--channel scheme, connecting in each step transitions in projectile and target. In double beta--decay, however, one is interested in a second order NME in one nucleus. Hence, the DSCE t--channel formulation must be changed to a s--channel representation. In \cite{Lenske:2021jnr} this transformation was achieved by a combination of several steps of angular momentum recoupling and contour integration techniques.
As a result, the mixed projectile--target first and second step response functions were separated into polarization tensors in projectile and target. For example, the DSCE transition $A \to B$ is described by rank--2 polarization tensors of the form
\be\label{eq:PiAB}
\Pi^{(AB)}_{(S_1S_2)SM}(\omega|\mathbf{p}_1,\mathbf{p}_2)=\sum_C
\frac{\left[F^{(BC)}_{S_2T}(\mathbf{p}_2)\otimes F^{(CA)}_{S_1T}(\mathbf{p}_1)\right]_{SM}}{\omega-(M_A-M^*_C)},
\ee
where $\mathbf{S}_{1,2}$ and $\mathbf{p}_{1,2}$ denote the spin and momentum transfer in the first and second excitation and $\mathbf{S}=\mathbf{S}_1+\mathbf{S}_2$ is the total spin transfer. The transition form factors, discussed in detail in Ref. \cite{Lenske:2021jnr}, are defined by
$F^{(CA)}_{S_1T}(\mathbf{p}_1)=\lan C| e^{i\mathbf{p}_1\cdot \mathbf{r}} \bm{\sigma}^{S_1} \bm{\tau} |A\ran$, $S_1=0,1$,  and $F^{(BC)}_{S_2T}(\mathbf{p}_2)$ is obtained accordingly. The
striking formal similarity to the NME of $2\nu 2\beta$ decay is obvious, as discussed in detail in section \ref{sec:DCE_DBD}.

\section{The Isotensor Majorana DCE Mechanism}\label{sec:MDCE}

\begin{figure}
\begin{center}
\sidecaption
\includegraphics[width = 4.6cm]{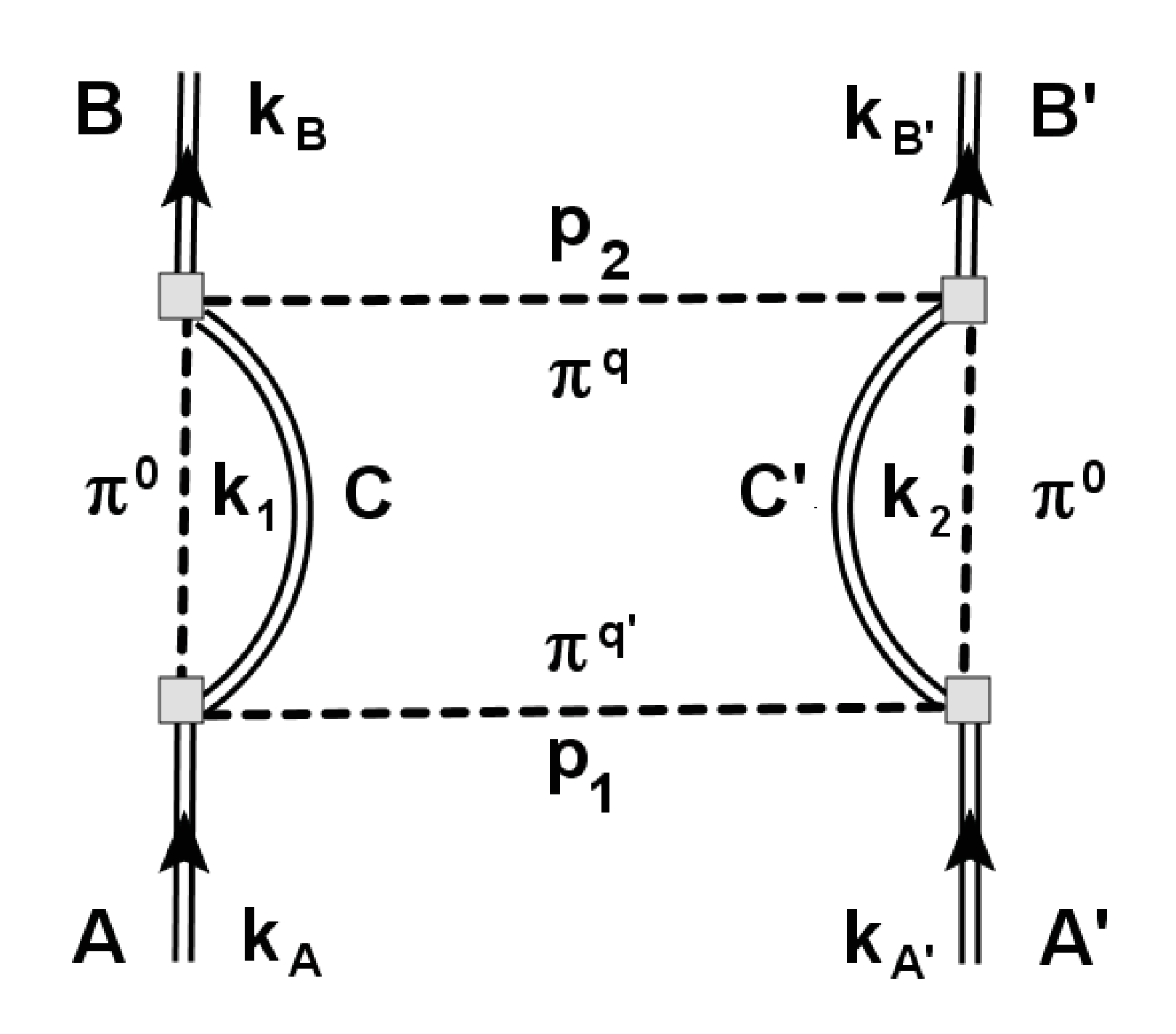}
\caption{The MDCE diagram for the reaction $A(Z,N)+A'(Z',N')\to B(Z\pm 2,N\mp 2)+B'(Z'\mp 2,N'\pm 2)$ and the involved momenta. The isovector pion--nucleon T--matrices are denoted by boxes. The two SCE events are correlated by $\pi^0$ exchange, making MDCE a two--nucleon process. The co--propagating core states are denoted by $C=C(Z\pm 1,N\mp 1)$ and $C'=C'(Z\mp 1,N\pm 1)$.  Charged pions $\pi^q$ and $\pi^q$, $q,q'=\pm 1$ are exchanged between the nuclei.  }
\label{fig:MDCE}
\end{center}
\end{figure}

A DCE reaction might also proceed as a single--step process if a rank--2 isotensor interaction would exist. However, hitherto searches have been unsuccessful both for elementary isotensor mesons \cite{Dover:1984zq,Wu:2003wf,Anikin:2005ur} as well as for interactions of that kind in nuclei \cite{Leonardi:1976zz}. Hence, most likely such an interaction is not realized as an elementary mode of its own right. However, if two nuclei are in close contact as in a peripheral ion--ion collision, an effective isotensor interaction can be generated dynamically. This case is investigated in the Majorana DCE (MDCE) reaction scenario \cite{Lenske:2019cex,Cappuzzello:2022ton}. As illustrated diagrammatically in Fig.\ref{fig:MDCE}, the MDCE mechanism relies dynamically on a pair of virtual ($\pi^\pm\to \pi^0 \to \pi^\mp$) reactions. The reacting nuclei interact by the t--channel exchange of charged pions, undergoing $\pi^\pm\to \pi^0$ reactions where each is inducing SCE--type
$A(Z,N)\to C(Z\pm 1,N\mp 1)$ transition. Propagating in the s--channel the neutral pion experiences a $\pi^0\to \pi^\pm$ reaction and excites a second SCE-transition $C(Z\pm 1,N\mp 1)\to B(Z\pm 2,N\mp 2)$.  As seen in Fig. \ref{fig:MDCE}, the off-shell ($\pi^+,\pi^-$) reaction in one nucleus is accompanied by a complementary ($\pi^-,\pi^+$) reaction in the other nucleus.

Theoretically, MDCE is described by the box diagrams of Fig.\ref{fig:MDCE}. Under nuclear structure aspects, the MDCE operator induces two--particle-two--hole ($p^2n^{-2}$) and ($n^2p^{-2}$) DCE transitions, respectively, in either of the interacting nuclei.

At first sight, DSCE and MDCE seem to be rather similar by the fact that projectile and target interact by charged meson exchange. The decisive different lies in the nuclear transition vertices. In the MDCE case they are determined by pion-nucleon dynamics which dynamically are completely different from NN--scattering.
At the energies relevant for the MDCE process the (off-shell) isovector pion--nucleon T-matrix $T_{\pi N}$  is described adequately by the operator structure \cite{Moorhouse:1969va,Johnson:1994na}
\be\label{eq:TpiN}
\mathcal{T}_{\pi N}(\mathbf{p},\mathbf{p}'|\bm{\sigma})= \left[T_0+\frac{1}{m^2_\pi}\left(T_1\mathbf{p}\cdot \mathbf{p}'+iT_2\bm{\sigma}\cdot(\mathbf{p}\times \mathbf{p}')\right)\right]\mathbf{T}_\pi\cdot \bm{\tau}_N .
\ee
The $T_0$ component is given by formation of $\pi N$ S--wave $N^*$ resonances $S_{2I2J}$ of negative parity with
isospin $I=\frac{1}{2},\frac{3}{2}$  and total angular momentum $J^\pi=\frac{1}{2}^-$.
The form factors  $T_{1,2}$ of the longitudinal and transversal parts originate from $P_{2I2J}$ configurations of positive parity with isospin as before but $J^\pi=\frac{1}{2}^+,\frac{3}{2}^+$. The most prominent P--wave resonances  are the Delta and the Roper resonances with spectroscopic notations $\Delta_{33}(1232)$ and $P_{11}(1440)$, respectively.  At the energies considered here, MDCE reactions take place off the pion-nucleon mass shell. Hence, pion-nucleon scattering must be described by methods allowing to extrapolate $T_{\pi N}$ into off--shell energy regions. That goal is achieved by appropriately modelling the pion self--energies with analytically given complex--valued form factors where parameters are adjusted to on--shell observables. In order to obtain converged below--threshold results, in practical calculations resonances up to the mass region of 2000~MeV must be taken into account.

The key elements for spectroscopy are the nuclear matrix elements $\mathcal{W}_{AB}$ and $\mathcal{W}_{A'B'}$, describing in Fig.\ref{fig:MDCE} the vertical left and right branches, respectively. They depend on the external momenta $\mathbf{p}_{1,2}$ attached to the charged pions and on the energy and momentum transfer to the nuclear SCE modes. For a DCE transition $A(Z,N)\to B(Z\pm 2,N\mp 2)$ the MDCE transition form factor is given by
\be\label{eq:Wij}
\mathcal{W}_{AB}(\mathbf{p}_1,\mathbf{p}_2)= -\sum_C \int \frac{d^3k}{(2\pi)^3}
\mathcal{M}_{BC}(\mathbf{p}_2,\mathbf{k})\frac{1}{k^2+m^2_\pi-\omega^2_{CA}}\mathcal{M}_{CA}(\mathbf{p}_1,\mathbf{k}),
\ee
where the summation extends over the intermediate SCE--type configuration $C$ and $\omega_{CA}=M^*_C-M_A$. The two charge--converting processes are described by SCE--type nuclear matrix elements, e.g. for the transition $C(Z\pm 1,N\mp 1)\to B(Z\pm 2,N\mp 2)$
\be\label{eq:Vci}
\mathcal{M}_{BC}(\mathbf{p},\mathbf{k})=\lan B|e^{i\mathbf{p}\cdot \mathbf{r}}
\mathcal{T}_{\pi N|\bm{\sigma}}(\mathbf{k},\mathbf{p})|C\ran ,
\ee
and $\mathcal{M}_{CA}$ is obtained correspondingly.

The pion rest mass $m_\pi\sim 139$~MeV defines a natural separation scale: Obviously, for energies $\omega_{CA}\ll m_\pi$, covering the intermediate states of primary importance for the reaction, the NME can be evaluated safely in closure approximation. Removing the explicit dependence on the intermediate states leads to the pion potentials \cite{Cappuzzello:2022ton}
\be \label{eq:PionPot}
\mathcal{U}_\pi(\mathbf{x}|\mathbf{p}_{1,2}\bm{\sigma}_{1,2})=-\int \frac{d^3k}{(2\pi)^3}
\mathcal{T}_{\pi N}(\mathbf{p}_2,\mathbf{k}|\bm{\sigma}_2)\frac{e^{i\mathbf{k}\cdot \mathbf{x}}}{k^2+m^2_{\pi^0}}
\mathcal{T}_{\pi N}(\mathbf{p}_1,\mathbf{k}|\bm{\sigma}_1),
\ee
where $\mathbf{x}=\mathbf{r}_1-\mathbf{r}_2$ is the distance between the two nucleons involved in the MDCE transition.
A closer inspection of Eq.\eqref{eq:PionPot} reveals that the pion potential  induces a correlation between the two nucleons participating in the transition. Thus, in either of the two nuclei we encounter a two--nucleon process, making the complete MDCE reaction in total to a four nucleon event.

The transition form factor, which defines the nuclear matrix element, is obtained as
\be \label{eq:WAB_Clos}
\mathcal{W}_{AB}(\mathbf{p}_1,\mathbf{p}_2)=
-\lan B|e^{-i\mathbf{p}_2\cdot \mathbf{r}_2}\mathcal{U}_\pi(\mathbf{x}|\mathbf{p}_{1,2}\bm{\sigma}_{1,2})e^{i\mathbf{p}_1\cdot \mathbf{r}_1}
\mathcal{I}_{2\pm 2}|A\ran ,
\ee
including the isotensor $\mathcal{I}_{2\pm 2}=\left[\bm{\tau}_1\otimes \bm{\tau}_2\right]_{2\pm 2}$. The MDCE pion potentials, acting in projectile and target, are two--body operators enforcing $n^{\mp 2}p^{\pm 2}$, i.e. $\Delta Z=\pm 2$, transitions, while conserving the total charge of the projectile--target system.

\begin{figure}
\begin{center}
\sidecaption
\includegraphics[width = 9cm]{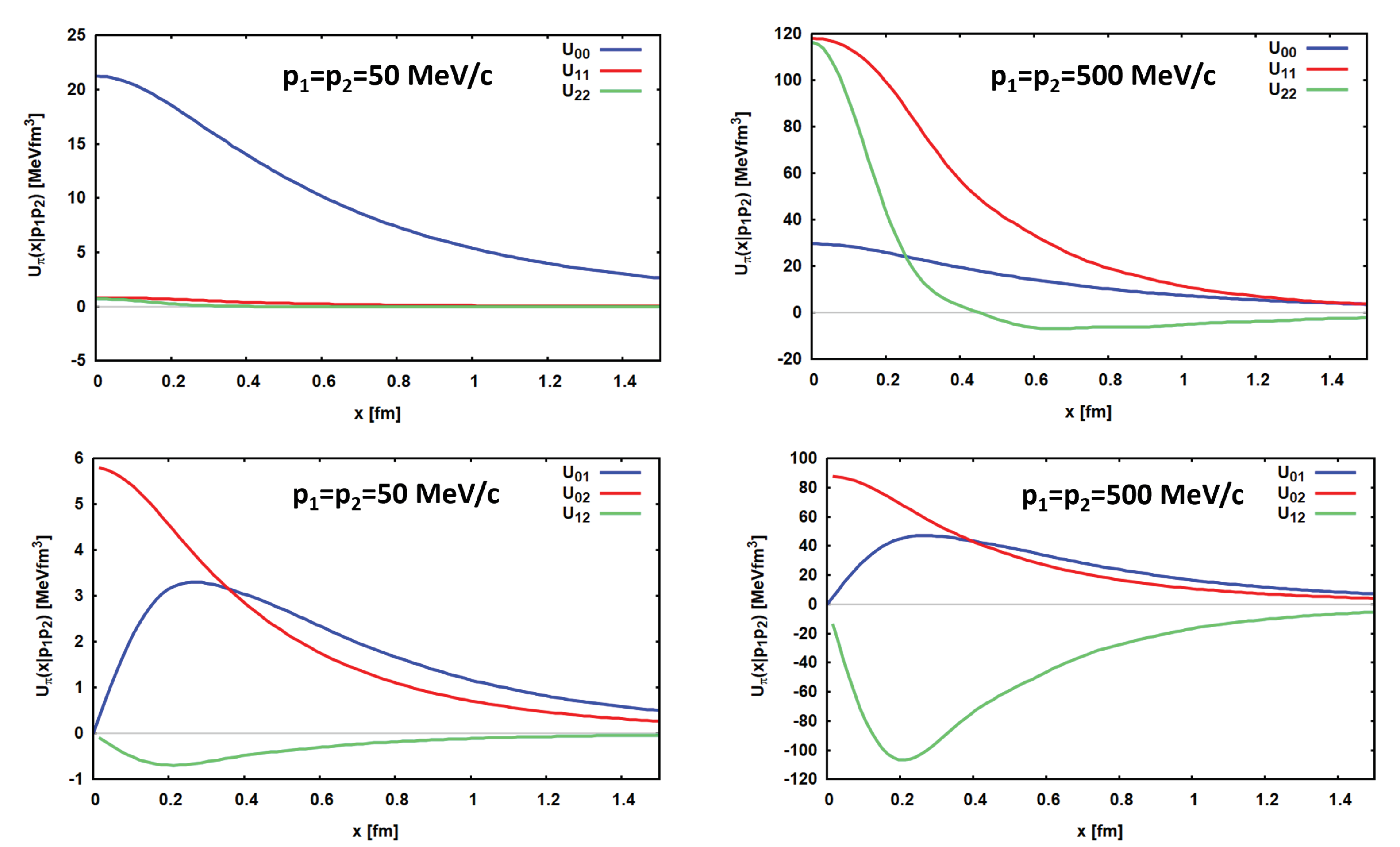}
\caption{The MDCE pion potentials for collinear off--shell $\mathbf{p}_1||\mathbf{p}_2$ momenta with $p_1=p_2=50$~MeV/c (left column) and $p_1=p_2=500$~MeV/c (right column), respectively.  }
\label{fig:PionPot}
\end{center}
\end{figure}

By Eq.\eqref{eq:TpiN}, the pion potential $\mathcal{U}_\pi(\mathbf{x})$ is found to consist of nine terms. For collinear momenta $\mathbf{p}_1 || \mathbf{p}_2$, they reduce to six independent scalar form factors, $U_{ij}(x|p_1,p_2)$, $i\leq j=0,1,2$.  Angular integrations can be carried out in closed form. The remaining the k--integrals, given by products of ordinary or spherical Bessel functions and 2$^{nd}$ kind Legendre functions times powers of $k$, have to be evaluated numerically. The momentum integrals are regularized by dipole form factors with cut-off $\Lambda = 2000$~MeV/c. Typical results for the pion potentials $U_{ij}$ obtained for $^{40}$Ca are shown in Fig.\ref{fig:PionPot}. At low momentum transfers ($p_{1,2}=50$~MeV/c), left column of Fig.\ref{fig:PionPot}, the diagonal potential $U_{ii}$ are dominated by the S--wave term $U_{00}\sim T^2_0$. With increasing momentum transfer the P--wave components ($i=j=1,2$) are enhanced, as seen in the right column of Fig.\ref{fig:PionPot} where results for $p_{i}=500$~MeV/c are displayed. The non--diagonal potentials $U_{ij}$, $i\neq j$, the mixed S--/P--wave potential $U_{02}$  prevails at small distances $x$, but $U_{01}$ and to some extent also $U_{12}$ become important at larger distances.

\begin{figure}
\begin{center}
\sidecaption
\includegraphics[width = 6.5cm]{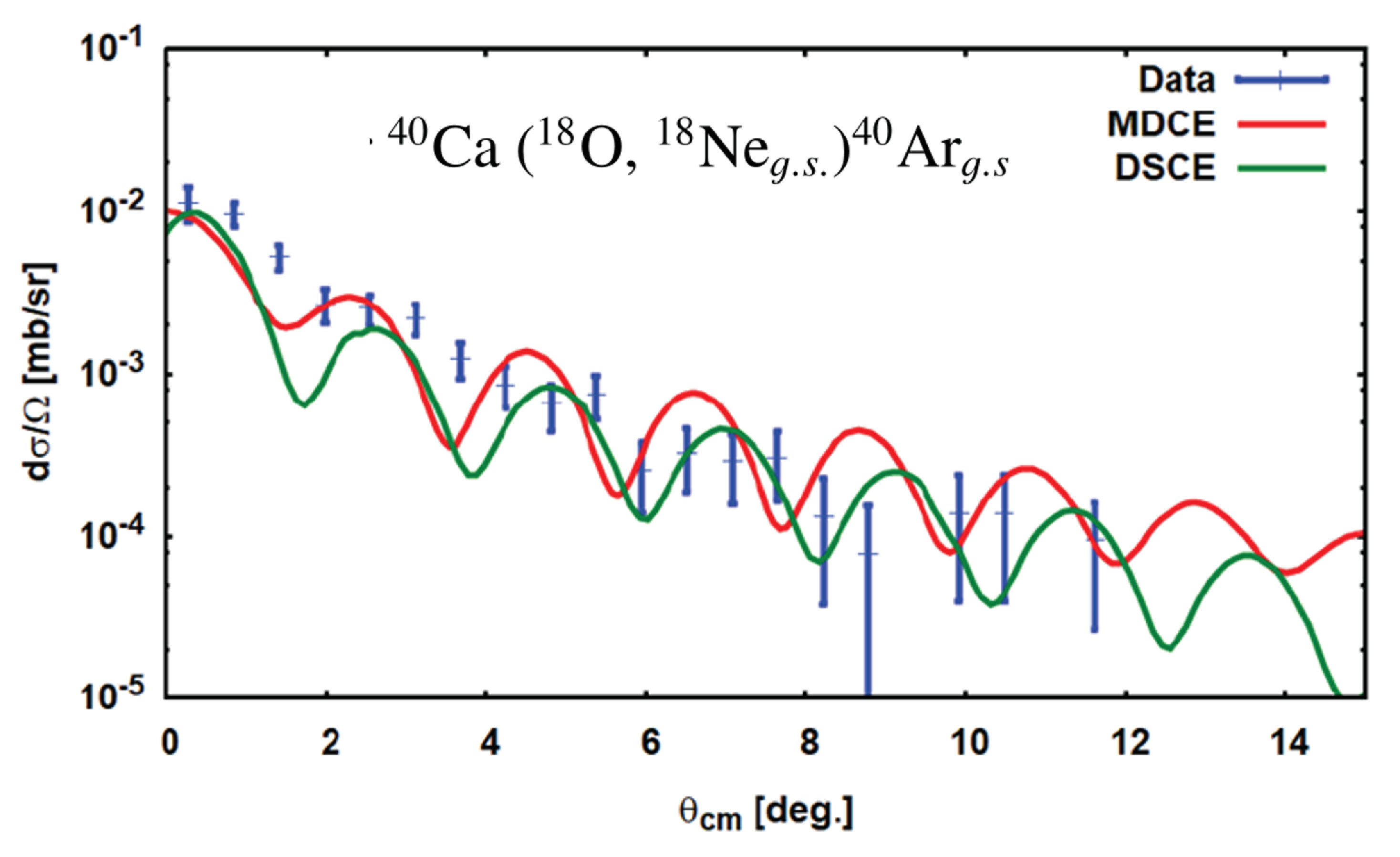}
\caption{DSCE and MDCE cross sections compared to the NUMEN data for $^{40}$Ca $(^{18}$O, $^{18}$Ne$_{g.s.})^{40}$Ar$_{g.s.}$ at $T_{lab}=275$~MeV \cite{Cappuzzello:2015ixp}. The MDCE cross section contains only contributions of the diagonal pion potentials $U_{ii}$ and is normalized to the forward angular distribution. Results are folded with the experimental angular resolution of $\pm 0.6^\circ$. }
\label{fig:MDCE_Data}
\end{center}
\end{figure}

This has important consequences for spectroscopic studies, because at forward angles preferentially double non--spin--flip Fermi--type  transitions will be excited by the scalar operator $\sim U_{00}$. The double spin--flip Gamow--Teller-type modes with an effective coupling constant $\sim U_{22}$ will contribute mainly at larger scattering angles. A different situation is encountered for the mixed potentials $U_{02}$ and $U_{12}$. They define the strengths of two--body operators containing only a single $\bm{\sigma}$--matrix. Contributions of such operators depend critically on the structure of the final DCE--state $B$ with respect to the parent state $A$. Suppose $|B\ran \sim [\Omega^\dag_{J_2}\otimes \Omega^\dag_{J_1}]_{J_B}|A\ran $ where $\Omega^\dag_{J_k}$ are state operators for which $|A\ran$ is the vacuum state. We consider three limiting cases:
\begin{itemize}
  \item Suppose that the state operators describe natural parity states, $\pi_{J_k}=(-)^{J_k}$. If $J_1=J_2=0^+$ are monopole excitations, as e.g. a double isobaric analogue resonance excitation, then only the spin--scalar parts $\sim U_{00}$, the longitudinal term $\sim U_{11}$, and the mixed scalar--longitudinal part $\sim U_{01}$ are relevant.
  \item If (at least) one of the states $J_k$ has non--zero angular momentum, then the spin--response will be non--zero. In this case, the  $U_{i2}$--potentials probe the content of natural parity spin--flip strengths induced by \textit{electric} spin--flip operators $\sim [Y_{J_k}(\hat{\mathbf{r}})\otimes \bm{\sigma}]_{J_k}$. If both $J_{1,2}>0$ that feature is probed also by the $U_{22}$--type double spin operators.
   \item If both states are unnatural parity states, $\pi_{J_k}=(-)^{J_k+1}$, then only the transversal spin--spin terms of strength $U_{22}$ will contribute.
\end{itemize}
These properties are maintained also for $p_1\neq p_2$ and in slightly modified form also for the general, non--collinear case.

The s--channel $\pi^0$ establishes in fact a rather tight two--nucleon correlation. The correlation lengths varies slightly with $p_i$ but rarely reaches 40\% of the root--mean-square range of pion exchange. Hence, the MDCE processes is of a pronounced short--ranged character. The correlated pair acts as a virtual, polarized pion dipole source. First preliminary results, neglecting the non--diagonal pion potentials and not taking into account the coherence of the MDCE and the DSCE reaction amplitudes, are shown in Fig.\ref{fig:MDCE_Data}. Because of the approximations the magnitude of the MDCE cross section is not yet fully fixed and was adjusted to the data at forward angles. Compared to the DSCE angular distribution the diffraction structure of the MDCE cross section is shifted towards smaller angles, filling the more pronounced minima of the DSCE component.  Remarkably, the angular region covered in Fig.\ref{fig:MDCE_Data} amounts to linear momentum transfers up to 500~MeV/c.

\begin{figure}
\begin{center}
\sidecaption
\includegraphics[width = 6.5cm]{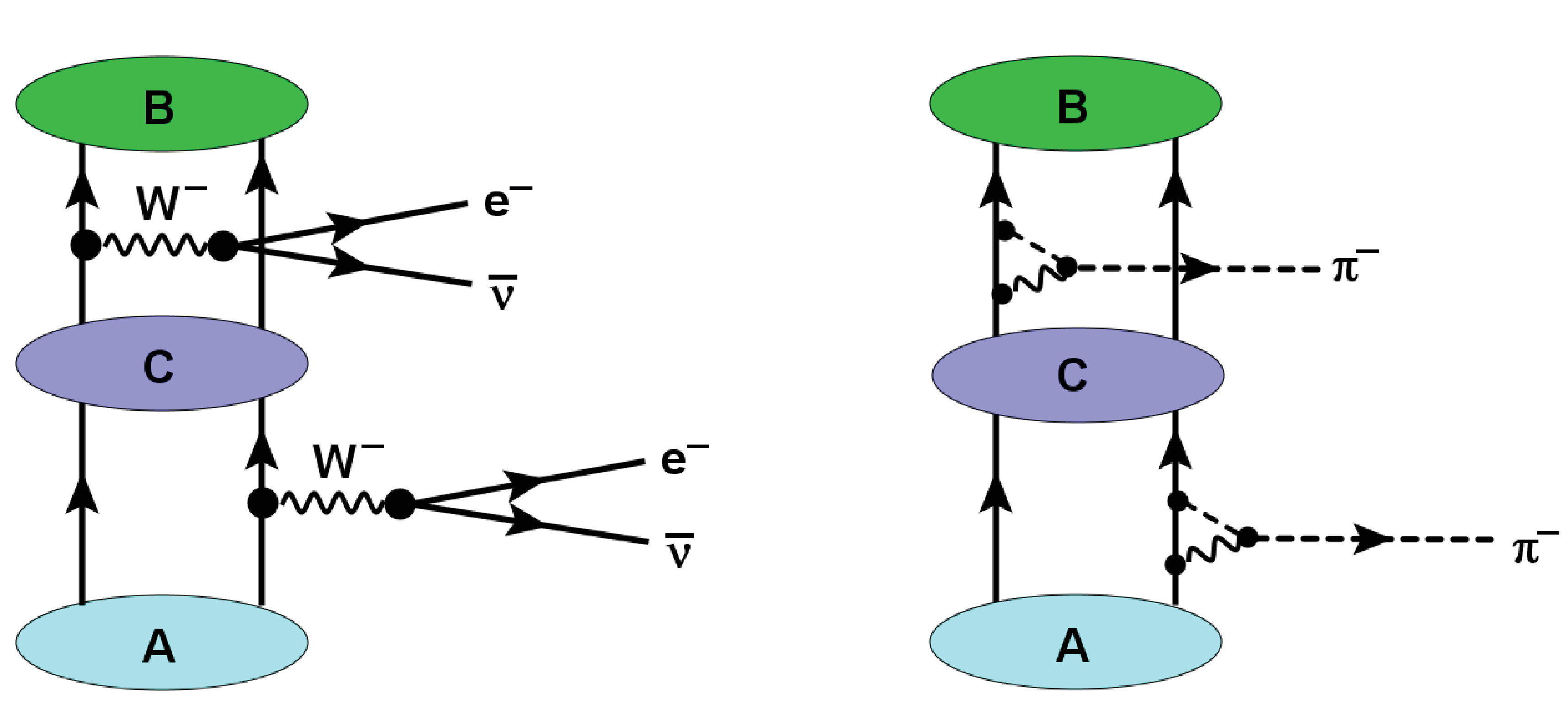}
\caption{Illustration of the $A(Z,N)\to B(Z+2,N-2)$ transition for two--neutrino DBD (left) and hadronic DSCE in one of the interacting nuclei (right) by typical elementary graphs of comparable topology. See text for further discussion. }
\label{fig:DSCE_NME}
\end{center}
\end{figure}

\section{Relation of Heavy Ion DCE Dynamics to Double Beta Decay}\label{sec:DCE_DBD}

The past experiences with SCE reactions, see e.g. the review article \cite{Lenske:2019cex}  hints at a close relation between DSCE processes and $2\nu 2\beta$ decay. Both are second order processes. Despite the large differences in dynamics, nuclear transitions are described by the same type of second order nuclear matrix elements given by the rank--2 polarization tensors derived in \cite{Lenske:2021jnr} and mentioned above. In Fig.\ref{fig:DSCE_NME} the formal similarity of a DSCE reaction and two--neutrino DBD also on the underlying dynamical levels is elucidated by comparing two typical diagrams. The DBD process, see e.g. \cite{Tomoda:1990rs,Ejiri:2019ezh}, is given by two half off--shell single beta decay (SBD) vertices where a highly virtual weak $W$ gauge boson materializes into a lepton--antilepton pair. On the hadronic level, the corresponding DSCE process proceeds by a pair of SCE--type vertices where a highly virtual vector--isovector rho--meson decays into a pair of off--shell pions, $\rho^{\pm}\to \pi^0 + \pi^\pm$, where the neutral pions are rescattered and recombined within the same nucleus while the charged pions will be transmitted to the reaction partner and induce there DSCE transitions of complementary isospin character. In practice, however, the displayed elementary weak and the strong interaction diagrams together with other relevant graphs are taken into account globally by effective vertex form factors. In the weak sector, axial and vector coupling constants $g_{A,V}\sim \mathcal{O}(1)$ define the leading order contributions. For hadrons, the meson--nucleon isovector coupling constants, e.g the pion--nucleon coupling $g_{\pi N}\sim \mathcal{O}(10)$, are playing the same role. In both cases, the vertices must be regularized by momentum--dependent form factors and, especially for hadrons, the scattering series have to summed to all orders which is achieved by solving a system of coupled Lippmann--Schwinger equations, see e.g. \cite{Lenske:2021bpk}. In other words, it will be very unlikely to measure selectively a single mechanism related to a specific Feynman diagram.

\begin{figure}
\begin{center}
\sidecaption
\includegraphics[width = 7cm]{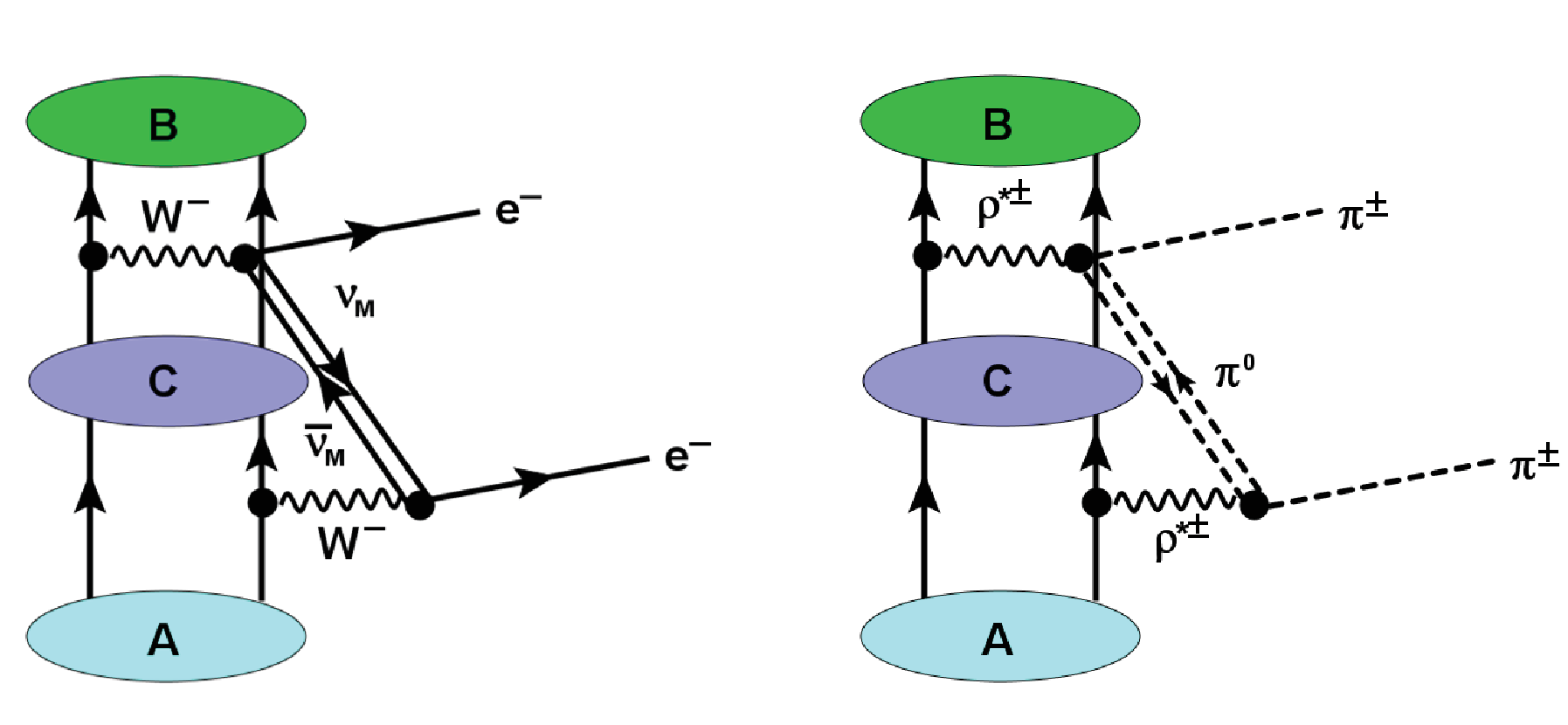}
\caption{Illustration by typical elementary graphs of comparable topology of the $A(Z,N)\to B(Z+2,N-2)$ transition for neutrinoless Majorana DBD (left) and the hadronic MDCE modes in one of the interacting nuclei (right). See text for further discussion. }
\label{fig:MDCE_NME}
\end{center}
\end{figure}

Also the similarities between MDBD and MDCE are understood the best by comparing typical interaction diagrams,  Fig.\ref{fig:MDCE_NME}. The MDBD process is initiated again by an off--shell charged gauge boson, decaying, however, in a $e^\pm$ and Majorana neutrinos
$\nu_M =\bar{\nu}_M$ \cite{Tomoda:1990rs,Ejiri:2019ezh}. That equality is the defining aspect of MDBD because it is essential for keeping the pair of Majorana neutrinos captured by exchange between the two vertices. Thus, they induce a correlation between the two nucleons taking part in MDBD.
Among the multitudes of processes of pion--nucleon interactions, ranging from point coupling and t--channel meson exchange to s--channel formation of $\pi N$ resonances, see e.g. \cite{Lenske:2018bgr}, in Fig. \ref{fig:MDCE_NME} we show the one of  diagrammatical form resembling closest the MDBD graph. The depicted (off--shell) rho--meson exchange process is closely related to the decay $\rho^\pm \to \pi^0 +\pi^\pm$. which on the mass--shell dominates the decay of a rho--meson. The displayed diagram corresponds can be understood as a rho--meson radiated off by the incoming pion there changing its charge and the rho--meson converts a neutron into a proton, hence creating a $pn^{-1}$ configuration. The intermediate, again highly virtual, $\pi^0$ converts upon the nuclear background into $\rho^+ +\pi^-$ resulting in a second $pn^{-1}$ transition induced by the rho--meson and an outgoing $\pi^-$. Two remarks are in place: firstly,  we emphasize that all processes occur far off their respective mass--shells; and secondly, the described \emph{rho--meson scenario} is not an unique source for final $[p^2n^{-2}]_b\otimes [n^2p^{-2}]_B$ projectile--target configurations but strongly competes with the resonance scenario by sequential $N^*n^{-1}\to pn^{-1}$ resonance formation--decay processes plus the proper kind of pions.

\section{Summary}\label{sec:Sum}
A fully microscopic approach to heavy ion collisional DCE reactions was presented, including sequential DSCE and direct MDCE contributions. A new spectroscopic property of DSCE reactions is that they are well suited to probe the little known rank--2 polarization tensors. They are generalizations of the polarization coefficients where the nuclear dipole polarizability is known best. The DSCE rank--2 nuclear polarization tensors occur also in $2\nu 2\beta$ decay. Moreover, they establish an interesting connection to double--gamma decay \cite{Soderstrom:2020iaz}.

The MDCE process relies on a hitherto unknown mechanism, namely a  dynamically induced rank--2 isotensor interaction.  In MDCE closure approximation, pion potentials and nuclear matrix elements are obtained close to those encountered in $0\nu 2\beta$ MDBD. The pion potentials include combinations of spin and momentum scalar parts, spin--scalar longitudinal and spin--vector transversal momentum--vector components, all attached to a rank--2 isotensor operator. This rich operator structure allows wide--spread spectroscopic studies, allowing a detailed tomography of the nuclear wave functions. However, experimentally and theoretically such studies are highly demanding because they require to observe, analyse, and interpreted energy--momentum distributions over large ranges.

\paragraph{\textbf{Acknowledgement:}} Support by 16$^{th}$ NRM, the dedication of the conference to my 70$^{th}$ birthday by the organizers, and the contributions of longtime friends, colleagues, and collaborators to the wonderful special session are gratefully acknowledged. This work was supported financially in part by DFG, grant Le439/16-2. J.-A. acknowledges that this work is based on research supported in part by Grant No. PID2020-114687GB-I00 funded by MCIN/AEI/10.13039/501100011033.

%

%
\end{document}